\documentclass[prl,twocolumn,amsmath,amssymb,superscriptaddress]{revtex4}

\usepackage{graphicx}

\begin{document}

\title{Conductance of p-n-p graphene structures with `air-bridge' top gates}
\author{R.~V.~Gorbachev}
\author{A.~S.~Mayorov}
\author{A.~K.~Savchenko}
\author{D.~W.~Horsell}
\affiliation{School of Physics, University of Exeter, Stocker Road, Exeter, EX4 4QL, U.K.}
\author{F. Guinea}
\affiliation{Instituto de Ciencia de Materiales de Madrid, CSIC, E28049 Madrid, Spain}

\begin{abstract}
We have fabricated graphene devices with a top gate separated from the graphene layer by an air gap---a design which does not decrease the mobility of charge carriers under the gate.  This gate is used to realise p-n-p structures where the conducting properties of chiral carriers are studied. The band profile of the structures is calculated taking into account the specifics of the graphene density of states and is used to find the resistance of the p-n junctions expected for chiral carriers. We show that ballistic p-n junctions have larger resistance than diffusive ones. This is caused by suppressed transmission of chiral carriers at angles away from the normal to the junction.
\end{abstract}

\maketitle

Graphene, a monolayer of carbon atoms, is a new two-dimensional system \cite{NovoselovS} with unusual properties: the gapless energy spectrum of electrons and holes is linear, and these carriers are chiral. It is the chirality that suppresses the backscattering of carriers \cite{Ando} and allows them to penetrate through potential barriers without reflection (the Klein paradox \cite{Katsnelson}). The theory of ballistic graphene p-n junctions \cite{Cheianov} predicts that the carriers propagate through the barrier without scattering if their angle of incidence is $0^\circ$ (with respect to the normal) and are partially reflected at other angles.  This angular selectivity of the carriers determines the resistance of a ballistic p-n junction.

To realise ballistic transport through a graphene p-n junction, the mean free path of carriers $l$ has to be larger than the characteristic length of the junction $2t$. A number of intriguing phenomena, such as small-field positive magnetoresistance \cite{Cheianov} and oscillating transmission probability \cite{Katsnelson}, can be observed in a ballistic p-n-p structure where the total length (two junctions plus the n-region) is smaller than $l$. Ballistic p-n and p-n-p structures can also be important for a number of potential applications: for example, graphene lenses \cite{Cheianov_l} and filter circuits \cite{Katsnelson} where the electrical current of chiral carriers is focused on or directed to a desired contact. Therefore, exploring the ways of producing high-mobility p-n-p graphene structures with large $l$ and understanding the mechanisms of carrier propagation through them is an important task.

Graphene flakes obtained by mechanical exfoliation \cite{NovoselovS} are conventionally deposited on a SiO$_2$/Si substrate where the conducting n-Si layer forms a (back) gate acting on the whole flake. Inversion of the type of carrier in a part of graphene flake has been recently achieved by using an additional (top) gate which is positioned above the graphene layer \cite{Goldhaber,Marcus,Kim,Morpurgo}. In the p-n junctions fabricated so far, the top gate rests on an insulating layer that can decrease the mobility of graphene carriers under it \cite{Goldhaber} and hinder the realisation of ballistic p-n and p-n-p structures. We avoid this problem by fabricating p-n-p graphene structures using suspended `air-bridge' top gates.  (Similar designs were used earlier in semiconductor nanostructures \cite{Yacoby_br,Girgis}.) The mobility of carriers under the top gate in such structures is the same as in the rest of the graphene layer.

The theory of ballistic graphene p-n junctions \cite{Cheianov} considers a `smooth' junction, $2k_Ft \gg 1$, where $k_F$ is the Fermi wave vector of the particles and $2t$ is the tunneling distance of the carriers. The transmission probability of such a junction as a function of the angle of incidence $\theta$ (with respect to the normal to the junction) is given by
\begin{equation}
w(\theta)=e^{-\pi \hbar v_F k_F^2 \sin^2{\theta}/F} = e^{-\pi \hbar v_F k_y^2/F}\;. \label{eqn:one}
\end{equation}
Here, $v_F$=10$^{6}$\,ms$^{-1}$ is the Fermi velocity of carriers and $F/e$ is the electric field in the barrier which is assumed to be constant over the tunneling distance so that the potential $\varphi(x)=Fx/e$.  The term $k_y$ is the component of the wavevector parallel to the junction,  $k_y=\pi n/W$, $n=1,2,\ldots$, where $W$ is the width of the barrier.
Therefore, the transmission of chiral carriers is restricted to the angles $\theta \leq \theta_c\simeq (F/\pi \hbar v_F k_F^2)^{1/2}$ or, equivalently, the transverse momentum values $k_y\leq (F/\pi \hbar v_F)^{1/2}$. The conductance of a ballistic junction is then \cite{Cheianov}
\begin{gather}\label{eqn:two}
R_{pn}^{-1}=\frac{4e^2}{h}\sum_n w(k_y)\\
\approx\frac{4e^2}{h}\frac{W k_F}{2\pi}\int_{-\frac{\pi}{2}}^{+\frac{\pi}{2}}w(\theta)\cos{\theta}\;\mathrm{d}\theta
\approx \frac{2e^2}{\pi h}W\sqrt{\frac{F}{\hbar v_F}}\;.\nonumber
\end{gather}

For the calculation of the resistance of a graphene p-n junction it is important to know accurately the electric field at the p-n boundary, which can be much larger than that in the rest of the barrier \cite{Fogler1}. This is caused by poor screening near the point of electro-neutrality (the Dirac point) where the density of carrier states decreases to zero. Therefore, to find the expected resistance of ballistic p-n junctions in our samples, we have computed the potential profile of the experimental p-n-p structures.  This is done for different combinations of the back and top gate voltages, taking into account the density of states in graphene which changes linearly with energy.

To fabricate ballistic p-n junctions, one positions the top gate close to graphene, which increases the field $F$ and decreases the tunneling distance $t(k_y,F)=\hbar v_F k_y/F$. In addition, the Fermi wavelength increases when the electron diffusively approaches the junction, up to the distance $\sim l/2$ from its centre when it passes through it ballistically. As a result, the length of the junction $2t$ can become comparable to the wavelength, and the assumption of \cite{Cheianov} of a smooth barrier not applicable. To examine the validity of this approximation for our structures, we have also found the transmission probability $w(\theta)$ by solving the Dirac equation numerically \cite{Guinea}, using the calculated potential profile $\varphi(x)$.

This calculated potential $\varphi(x)$ allows us to determine the resistance of the p-n-p structure expected for diffusive propagation of carriers (without taking into account their chirality) and to compare the result with experiment. The required resistivity $\rho(\varphi)$ of the sample at different Fermi energies is found from the resistance as a function of back-gate voltage $V_{bg}$ (at top-gate voltage $V_{tg}=0$).

Here, we present the results for three samples with different mobilities. The sample with the smallest mean free path has a resistance in the p-n-p regime which is in agreement with the diffusive model, while the higher-mobility samples (with ballistic transport through the p-n junctions) show an enhanced value of the resistance, which agrees with the result we expect for the chiral carriers.

\begin{figure}[htb]
\includegraphics[width=.98\columnwidth]{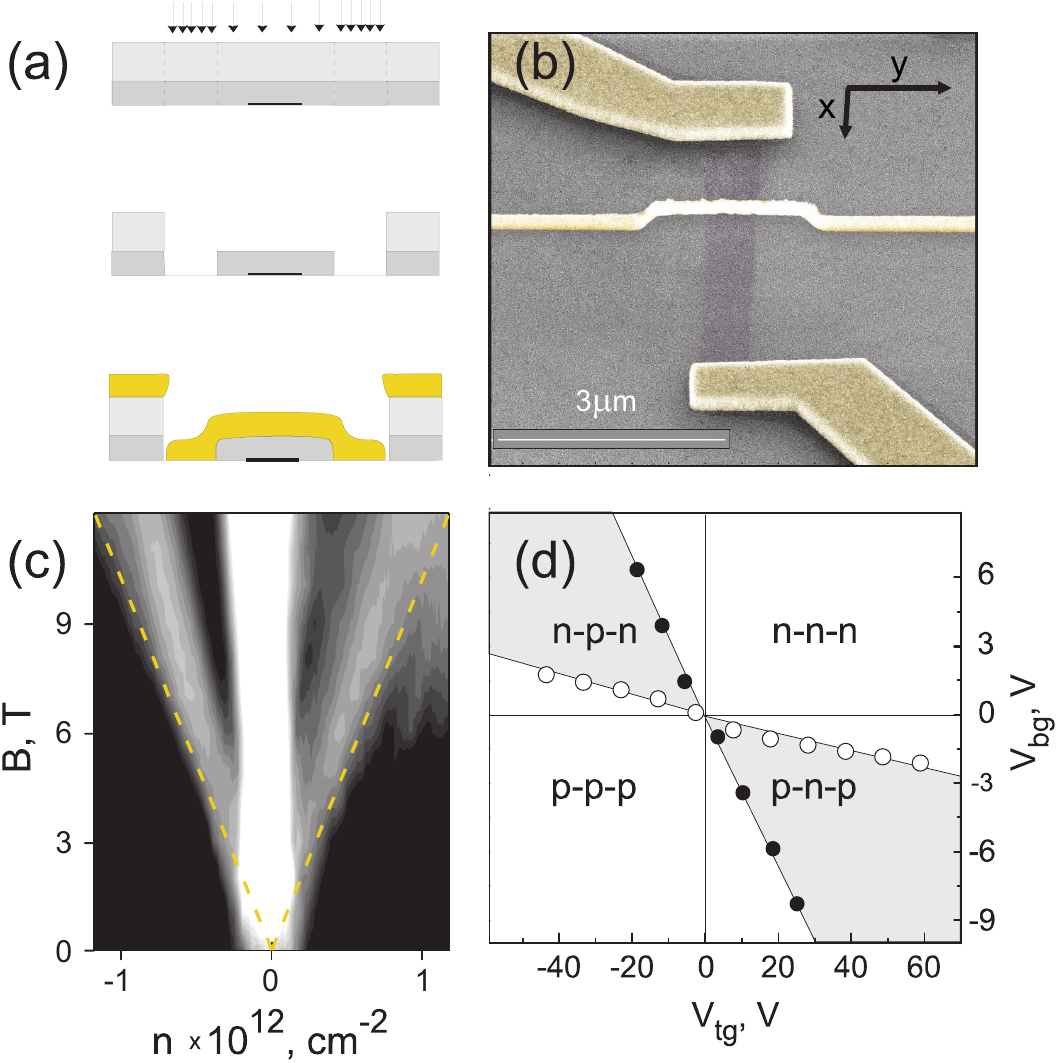}
\caption{(a) Three stages of the air-bridge fabrication: electron beam lithography with two exposure doses; development, and deposition of the metal film. (b) A false-colour SEM image of a graphene flake with a metal air-bridge gate, tilted at 45$^{\circ}$. (c) Grey-scale of the positions of the maxima in $R$ as a function of carrier density and magnetic field, with dotted lines corresponding to the shifts of the lowest Landau levels expected for single-layer graphene. (d) Positions of the resistance maxima at different $V_{bg}, V_{tg}$, with different regimes of the device operation indicated, sample S1. }\label{fig1}
\end{figure}

To fabricate the p-n-p structures, we have chosen graphene flakes of rectangular geometry on SiO$_2$/Si substrates with a 300\,nm oxide layer. The samples have the following dimensions, in $\mu$m: $L$=5, $W$=0.24 (sample S1); $L$=4.3, $W$=0.6 (sample S2) and $L$=1.45, $W$=0.15 (sample S3). The mobility of these samples outside the region of electro-neutrality (at a carrier density of $3\times10^{11}$\,cm$^{-2}$)  is $13$, $11$ and $6\; \times10^3$\,cm$^2$V$^{-1}$s$^{-1}$, respectively. The procedure of the top gate fabrication is illustrated in Fig.~\ref{fig1}(a).  Two layers of PMMA with different molecular weights are spun on the flake: a soft resist (495K) on top of a hard resist (950K). They were then patterned using 10\,kV e-beam lithography (to achieve larger undercut in the top PMMA layer). Two different exposure doses were used in the areas of the span and pillars of the bridge, while the area outside the bridge was not exposed. The dose in the span is just enough to expose the soft resist, but too small to affect the underlying hard layer. Both layers are exposed at a larger dose in the areas of the pillars (and contacts). The structures are then developed and covered with 5/250 nm of Cr/Au.  The `lift off' removes PMMA leaving the bridge with a span up to $2\, \mathrm{\mu m}$ supported by two pillars. Figure~\ref{fig1}(b) shows an SEM image of sample S2 with a bridge top gate and two Ohmic contacts. The mean free path in our samples $l\approx45-100$\,nm and the distance between the top gate and the flakes is 130--210\,nm. In the attempt to produce a p-n-p structure with ballistic properties, the top gate is made short in the direction of the current flow: 100--170\,nm.

Two-terminal measurements of $R(V_{bg})$ in quantising magnetic fields have confirmed that we are dealing with single-layer graphene. The grey-scale plot in Fig.~\ref{fig1}(c) shows the shift with magnetic field of two resistance peaks corresponding to the first electron and hole Landau levels. In this experiment, the carrier concentration was varied by the back gate voltage: $n/V_{bg}=7.2\times10^{10}\,\mathrm{cm^{-2}V^{-1}}$  (the relation from the known capacitance of the structure with a 300\,nm SiO$_2$ layer). A positive (negative) $V_{bg}$ induces electrons (holes) in the graphene layer, and the 0th Landau level (the bright vertical line) corresponds to the Dirac point. The dotted lines show the position of the 1st Landau level of electrons and holes in accordance with the filling factor $\nu=4eBn/h=\pm1$ \cite{NovoselovP, Haldane}.

Figure~\ref{fig1}(d) shows that, by changing the combination of the voltages on the back gate and top gate, one can get different regimes of the operation of the device, with both p-n-p and n-p-n modes available. Different regions are separated by the resistance peaks corresponding to the Dirac points under the top gate (steep line) and back gate (almost horizontal line).  The slope $\mathrm{d}V_{bg}/\mathrm{d}V_{tg}$ of the steep line (from 0.24 to 0.4 in our samples) gives the efficiency of top-gate control with respect to that of the back gate.

\begin{figure}[htb]
\includegraphics[width=.98\columnwidth]{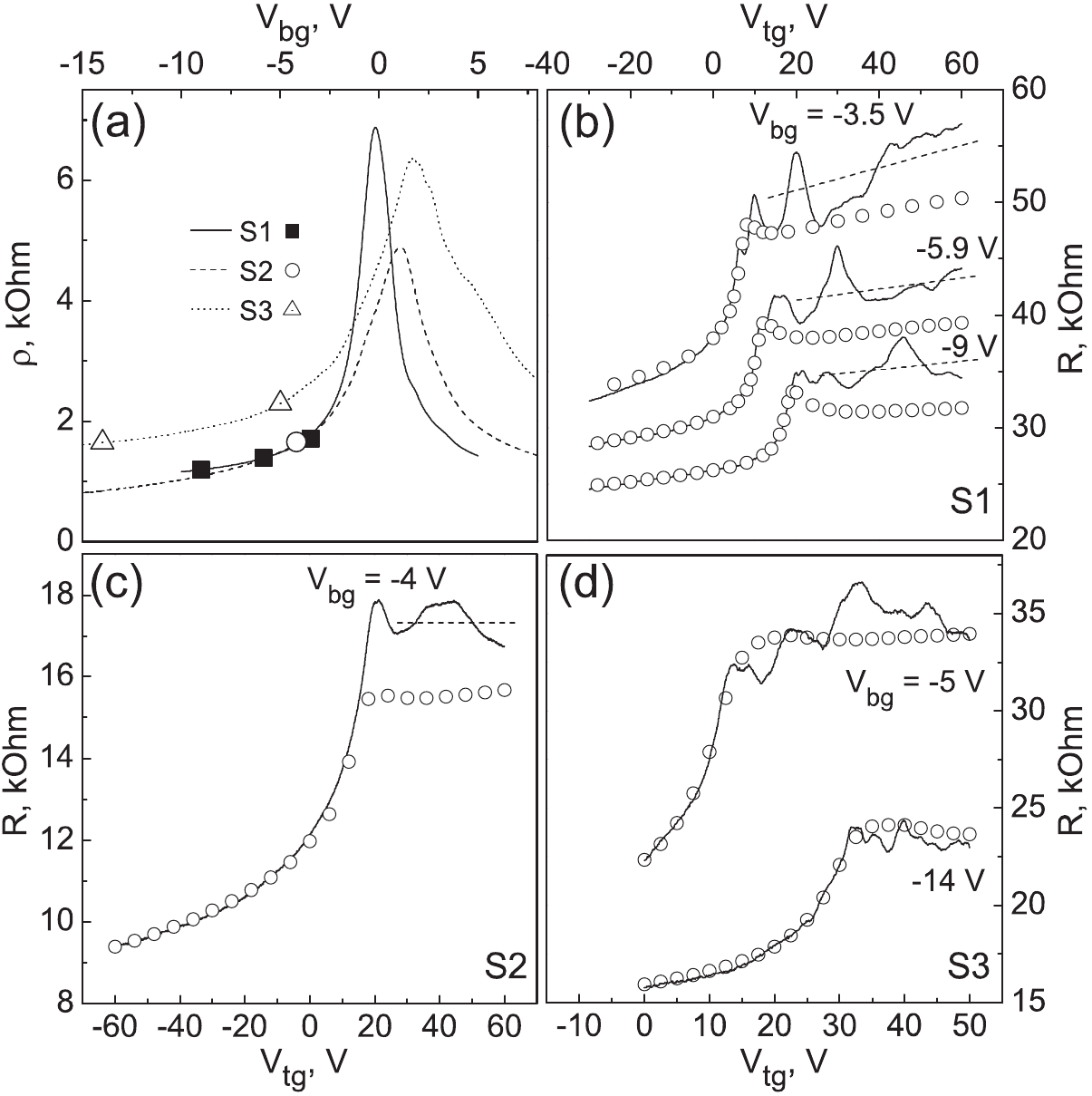}
\caption{(a) Resistivity of the three samples as a function of the back-gate voltage, at $V_{tg}=0$.  Points indicate the fixed values of $V_{bg}$ where the top-gate voltage was swept to produce p-n-p junctions. (b) The resistance of sample S1 as a function of top-gate voltage at different $V_{bg}$. (c,d) The resistance as a function of top-gate voltage at different $V_{bg}$ of samples S2 and S3, respectively. Points show the results of the calculations of the expected resistance assuming diffusive transport of carriers. (Dashed lines in b,c are guides to the eye.)}\label{fig2}
\end{figure}

\begin{figure}[htb]
\includegraphics[width=.98\columnwidth]{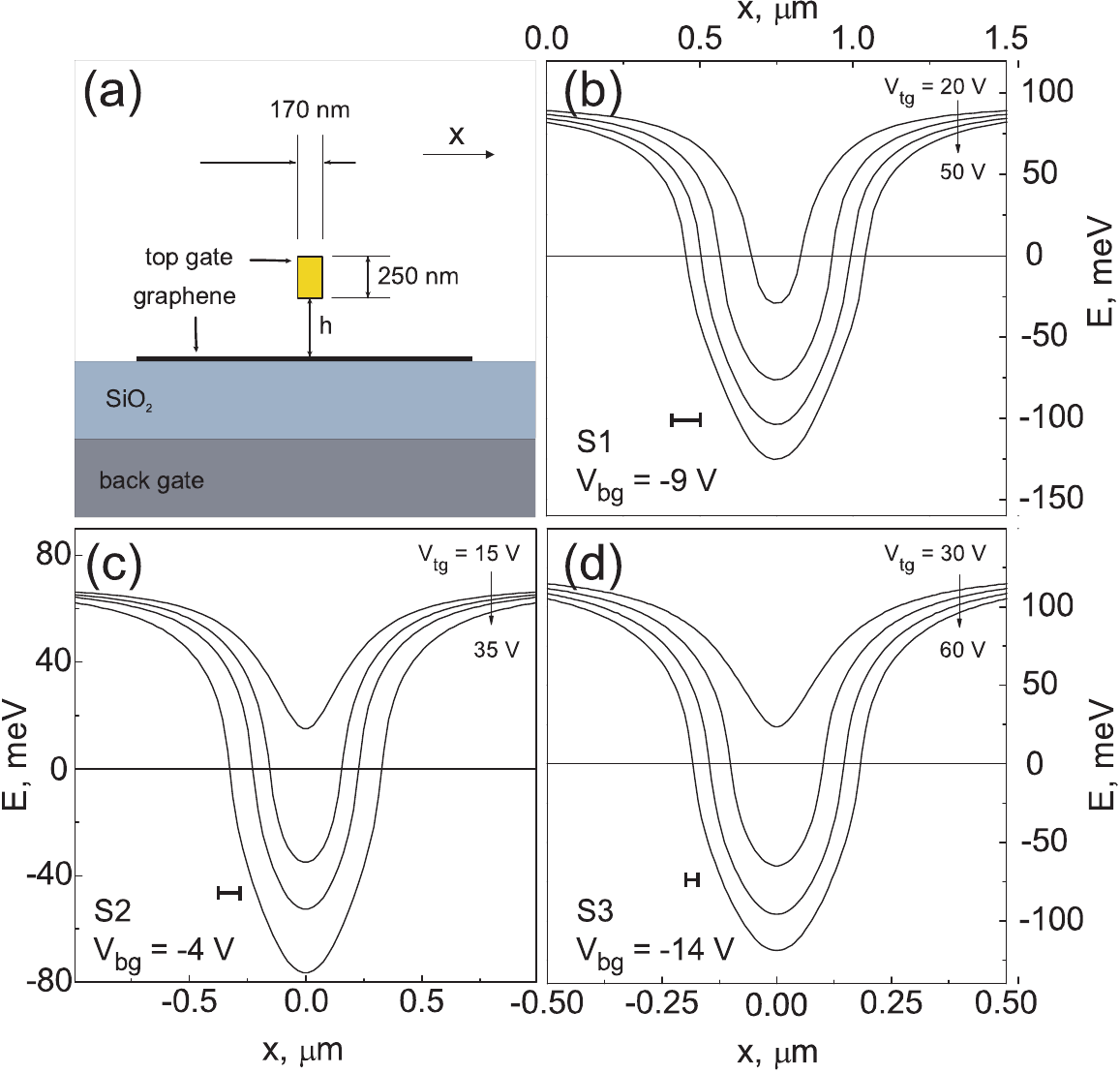}
\caption{(a) Geometry of the top-gated structure used in the calculations of the potential profile along the flake, sample S1.  (b,c,d) Potential profile of the top-gated samples S1, S2 and S3 along the barrier at different $V_{tg}$ for a fixed $V_{bg}$. The curves correspond to the position of the Dirac point and zero is the Fermi level. The bars show the mean free path $l$.}\label{fig3}
\end{figure}

Figure~\ref{fig2}(a) shows $R(V_{bg})$ at $V_{tg}=0$, where the peak corresponds to the Dirac point. Figure~\ref{fig2}(b-d) shows how the resistance of the samples changes (for fixed $V_{bg}$ values shown by points in Fig.~\ref{fig2}(a)) when a top-gate voltage is applied. When the main part of the sample outside the top gate is p-type, applying a negative $V_{tg}$ decreases the resistance due to the increase of the hole density under the top gate. Applying a positive $V_{tg}$ increases the resistance of the samples, first because of the depletion of electrons under the top gate and then because of inversion of the sign of carriers and formation of a p-n-p structure.

At the onset of the formation of the p-n junction, the resistance shows reproducible oscillations as a function of $V_{tg}$.  They survive at high temperatures, which strongly suggests that they can be due to the oscillations of the transmission coefficient caused by interference of chiral carriers \cite{Katsnelson} within a ballistic p-n-p structure. However, this effect has to be separated from mesoscopic fluctuations of resistance \cite{Gorbachev} that can be enhanced in the Dirac points of the p-n junctions. (We will discuss the separation of these effects elsewhere \cite{Mayorov}.)  Here, we consider the average values of the resistance and present the results obtained in a higher temperature range ($T$=50--77\,K) where the effect of the mesoscopic fluctuations is less important.

\begin{figure}[htb]
\includegraphics[width=.98\columnwidth]{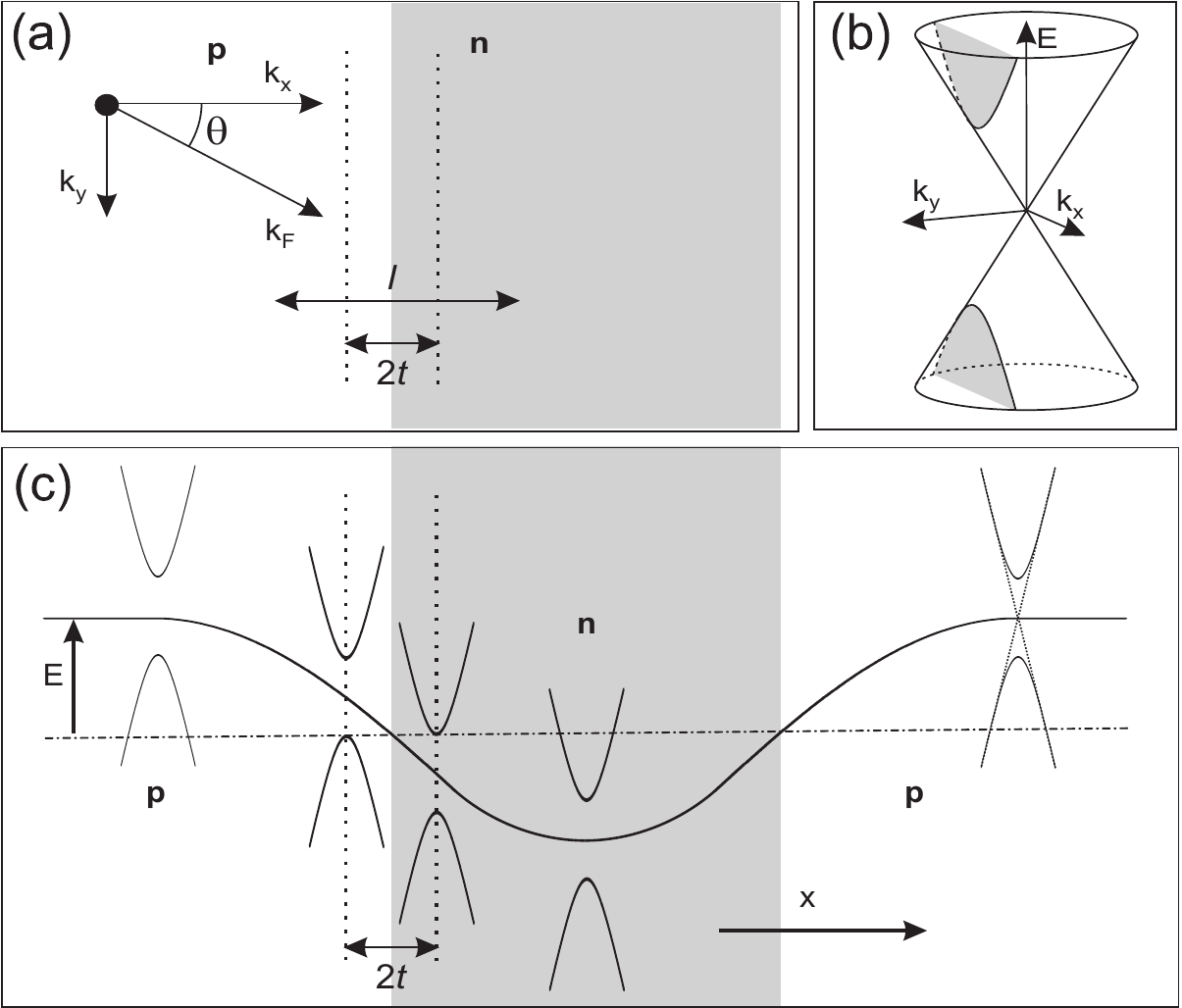}
\caption{(a)  Momentum of the electron approaching the junction at an angle $\theta$. (b) The gap in the spectrum $E(k_x)$ (highlighted) at $\theta \neq 0$. (c) Band-structure profile along the length of the p-n-p structure. The value of the gap determines the tunneling length $2t(\theta)$.}\label{fig4}
\end{figure}

The band-structure profile along the p-n-p structure at different $V_{tg}$ is calculated numerically for the geometry shown in Fig.~\ref{fig3}(a), by solving the two-dimensional Laplace equation with the potentials of the two gates as boundary conditions. The presence of the flake is included as an additional boundary condition on the jump of the normal component of the displacement field at the flake: $\Delta D_n=en(x)$, where $n(x)$ is carrier concentration along the sample. Taking the Fermi level as zero and using the linear energy dependence of the density of states in graphene, $\nu(E)=2E/\pi\hbar^2v_F^2$, one gets the relation between the carrier concentration along the junction $n(x)$ and the electrostatic potential: \mbox{$n(x)=\textrm{sgn}(\varphi)e^2\varphi^2(x)/\pi (\hbar v_F)^2$}, where sgn$(\varphi)$ reflects the fact that carriers can be both electrons and holes depending on the position of the Dirac point with respect to the Fermi level. Examples of such calculations for the three samples are shown in Fig.~\ref{fig3}(b-d), where indeed one can see a rapid increase of the electric field at the boundaries between p- and n-regions considered in \cite{Fogler1}: $F=(0.8-2.4)\times10^6$\,eV/m.

We use the `calibration' curves $R(V_{bg})$ in Fig.~\ref{fig2}(a) to find the resistivity of the flake at different Fermi energies (different electrostatic potentials if the Fermi level is taken as zero). We used the relation between $V_{bg}$ and the electrostatic potential $\varphi$ known from the capacitance between the flake and the back gate: $\varphi(\textrm{mV})=31\sqrt{V_{bg}(\textrm{V})}$. With the calculated distribution of the potential along the structure $\varphi(x)$, the integration of the resistivity $\rho(\varphi)$ gives the resistance expected for diffusive propagation of carriers: $R=(1/W)\int\rho(x)\,\mathrm{d}x$.

In the range of $V_{tg}$ corresponding to accumulation (negative $V_{tg}$) and depletion (small positive $V_{tg}$) under the top gate, the resistance is well-described by the diffusive model, Fig.~\ref{fig2}(b-d). One adjustable parameter, the distance $h$ between the top gate and the graphene flake, was used in plotting the calculated values: $h=$140, 210 and 130\,nm for samples S1, S2 and S3, respectively. The obtained values are close to those expected from the fabrication process and agree with observed efficiency of the  top gate, Fig.~\ref{fig2}(d).   With larger positive $V_{tg}$ and formation of the p-n-p structures, samples S1 and S2 show significantly larger values of the resistance than expected from the diffusive model: $\Delta R \simeq$4 and 2\,kOhm, respectively.  However, the narrowest sample S3 with the lowest mobility shows agreement with the diffusive model in the whole range of $V_{tg}$, Fig.~\ref{fig2}(d).

To explain these observations, we find the characteristic thickness of the p-n junctions in the three samples and compare it with the mean free path $l$. According to \cite{Cheianov}, the reason for the enhanced resistance of a junction is the decrease of the transmission when the electron approaches the junction at an angle $\theta \neq 0$, Fig.~\ref{fig4}(a). Conservation of the parallel component of the momentum $k_y$ produces a gap in the energy spectrum $E(k_x)$ for the motion across the junction, Fig.~\ref{fig4}(b). The distance $2t$ is then defined as the classically inaccessible region which requires electrons to tunnel along it, Fig.~\ref{fig4}(c): $t=\hbar v_F k_F\sin\theta/F$.  The critical angle for carrier transmission in the three samples varies in the range $\theta_c=20-30^\circ$, assuming the length of the ballistic p-n junction to be $l$ and taking the $k_F$-value at a point $x=-l/2$ from the barrier, Fig.~\ref{fig4}(a).  As the tunneling distance $2t$ depends on the angle of incidence, we take  for a typical value of the barrier thickness $2t(2\theta_c)\simeq$40\,nm in our samples.

The mean free path $l$ has been found using $R(V_{bg})$ of a uniform sample at $V_{tg}=0$, Fig.~\ref{fig2}(a), and the relation $\sigma=2e^2(k_Fl)/h$. The value of $l$ weakly depends on $V_{bg}$, and when extrapolated to the Dirac point ($V_{bg}=0$ for an undoped sample) gives $l\simeq$ 100, 75 and 45\,nm, respectively, for samples S1, S2 and S3. Comparing the tunnelling length with the mean free path shows that the p-n junctions in S1 and S2 are ballistic ($l\gg2t$), while in S3 they are less ballistic ($l\sim2t$). This can explain the agreement of the resistance of S3 with the result of the diffusive model in Fig.~\ref{fig2}(d).

To find the expected resistance $R_{pn}$ of ballistic p-n junction in samples S1 and S2 and compare it with the observed difference $\Delta R$ in Fig.~\ref{fig2}, we first assume a smooth potential barrier, $2k_Ft \gg 1$, and by using the calculated value of electric field $F$ we get the tunneling probability $w(\theta)$ from Eq.~\ref{eqn:one}. Equation~\ref{eqn:two} is then used to obtain the resistance of the ballistic p-n junction. We have found that using summation rather than integration is more appropriate in our case, as samples S1 and S2 have less than 12 modes (the narrowest sample S3 has only three modes).  The value of the Fermi momentum $k_F$ in these calculations is taken at a distance $l/2$ from the barrier using the values of the mean free path found above; however, the result for $R_{pn}$  hardly changes if the value of $l$ is varied by two times either way. This is clear as the tunneling probability $w(\theta)$ in Eq.~\ref{eqn:one} depends only on $k_y$ which takes specific, quantised values $k_y=\pi n/W$. The obtained values are $R_{pn}=$5 and 2\,kOhm for samples S1 (at $V_{bg}=-9$\,V, $V_{tg}=40$\,V) and S2 (at $V_{bg}=-4$\,V, $V_{tg}=30$\,V).

Taking into account the Fermi wavelength at the distance $l/2$ from the barrier, we see that $2k_Ft \simeq$2 for the three samples. To examine the applicability of a smooth-barrier approximation for this (not too large) value of $2k_Ft$, we have calculated $w(\theta)$ directly using numerical methods \cite{Guinea} and compared the result with that obtained from  Eq.~\ref{eqn:one}. It shows less than 5$\%$ difference from the value of $R_{pn}$ calculated above and a significantly larger resistance than the one expected for a sharp, rectangular barrier where $w(\theta)= \cos^2 \theta$ \cite{Cheianov}.

In experiment, it is not the resistance of an individual ballistic p-n junction which is measured but the resistance of the whole p-n-p structure. It can be different depending on whether its middle, n-region is long or short compared with $l$ (i.e., diffusive or ballistic). For a diffusive n-region with three independent contributions (two junctions and middle region) $R_{pnp}\geq 2 R_{pn}$, while for a ballistic n-region, $R_{pnp} \simeq R_{pn}$ \cite{Cheianov}. The resistance of a ballistic p-n-p structure should not increase with addition of another junction as the electrons approaching the second junction have already been selected by the first junction within the critical angle $\theta_c$. Therefore,  they all will have high transmission probability $w(\theta)$ going through the second junction.

Figure~\ref{fig2}(b,c) shows clearly that the resistance of S1 and S2 is larger than that expected in the diffusive model by $\Delta R$, because of the ballistic transport of chiral carriers through two p-n junctions. To find their resistance, we assume that they are independent; that is, the n-region is diffusive. Then the observed difference $\Delta R=2(R_{pn}-R_{pn}^D)$, where $R_{pn}^D$ is the resistance of the diffusive p-n junction on the length $l$ which was taken into account in the diffusive-model calculation shown in Fig.~\ref{fig2}. With the values $l=$100 and 75\,nm, one finds that $R_{pn}^D=$2 and 0.6\,kOhm for samples S1 and S2, respectively. This gives the corresponding resistance of the ballistic p-n junction $R_{pn}=$4 and 1.6\,kOhm, which is close to the expected values of 5 and 2\,kOhm. (Even better agreement, within 10\%, is achieved if another quantisation rule for graphene is used \cite{Tworzydlo}: $k_y=\pi(n+1/2)/W$, $n=0,1,2,\ldots$) The assumption of the diffusive nature of the n-region at large  $V_{tg}$ is confirmed by Fig.~\ref{fig3}, where the whole p-n-p region is seen to be larger than the mean free path. However, near the onset of the p-n junctions, at small $V_{tg}$, the p-n-p region is much shorter and can be fully ballistic.

In conclusion, we have fabricated p-n-p and n-p-n graphene structures using non-perturbative `air-bridge' top gates.  The chiral nature of charge carriers in graphene has been directly demonstrated by detecting an increase of the resistance of p-n junctions caused by their selective effect on the propagation of chiral particles. Our detailed analysis shows that individual p-n junctions are ballistic and that a ballistic p-n-p structure can be realised using this fabrication method.

We are grateful to V.~Cheianov, V.~Fal'ko, M.~Fogler, M.~Katsnelson and E.~McCann for stimulating discussions and EPSRC for funding. We also wish to thank P. Vukusic for generously allowing us access to his optical facility.

\end{document}